\documentclass[letterpaper]{article} 
\usepackage{aaai2026}  
\usepackage{times}  
\usepackage{helvet}  
\usepackage{courier}  
\usepackage[hyphens]{url}  
\usepackage{graphicx} 
\usepackage{natbib}  
\usepackage{caption} 
\usepackage{algorithm}
\usepackage{algorithmic}
\usepackage{booktabs}
\usepackage{siunitx} 
\usepackage{amsmath, amssymb, mathtools}
\usepackage{bm} 
\usepackage{booktabs}  
\usepackage{siunitx}   
\usepackage{newfloat}
\usepackage{listings}
\usepackage{graphicx}

\DeclareCaptionStyle{ruled}{labelfont=normalfont,labelsep=colon,strut=off} 
\floatstyle{ruled}
\newfloat{listing}{tb}{lst}{}
\floatname{listing}{Listing}
\title{RefAdGen: High-Fidelity Advertising Image Generation} 
\author{
Yiyun Chen\textsuperscript{\rm 1},
        Weikai Yang\textsuperscript{\rm 2}
}
\affiliations{

 \textsuperscript{\rm 1}The Hong Kong University of Science and Technology (Guangzhou)\\
    ychen780@connect.hkust-gz.edu.cn,
    weikaiyang@hkust-gz.edu.cn
}

\usepackage{bibentry}
\newcommand{\data}{AdProd-100K}
\newcommand{\model}{RefAdGen}

\begin{document}

\maketitle

\begin{abstract}
The rapid advancement of Artificial Intel-
ligence Generated Content (AIGC) techniques has unlocked opportunities in generating diverse and compelling advertising images based on referenced product images and textual scene descriptions.
This capability substantially reduces human labor and production costs in traditional marketing workflows.
However, existing AIGC techniques either demand extensive fine-tuning for each referenced image to achieve high fidelity, or they struggle to maintain fidelity across diverse products, making them impractical for e-commerce and marketing industries.
To tackle this limitation, we first construct \data{}, a large-scale advertising image generation dataset.
A key innovation in its construction is our dual data augmentation strategy, which fosters robust, 3D-aware representations crucial for realistic and high-fidelity image synthesis.
Leveraging this dataset, we propose \model{}, a generation framework that achieves high fidelity through a decoupled design. The framework enforces precise spatial control by injecting a product mask at the U-Net input, and employs an efficient Attention Fusion Module (AFM) to integrate product features. This design effectively resolves the fidelity-efficiency dilemma present in existing methods.
Extensive experiments demonstrate that \model{} achieves state-of-the-art performance, showcasing robust generalization by maintaining high fidelity and remarkable visual results for both unseen products and challenging real-world, in-the-wild images. This offers a scalable and cost-effective alternative to traditional workflows. 
Code and datasets are publicly available at {https://github.com/Anonymous-Name-139/RefAdgen}.

\end{abstract}

\section{1. Introduction}

In the fast-paced digital marketing and e-commerce landscape, there is a growing demand for quickly generating visually engaging advertising images~\cite{marwan2024impact}.
Traditional content creation relies on costly photoshoots and manual design, creating bottlenecks that hinder both speed and scale~\cite{cui2025exploring, adepoju2024automated}. While Artificial Intelligence Generated Content (AIGC), particularly diffusion models~\cite{ho2020denoising, rombach2022highresolution}, promises to revolutionize this landscape, a core challenge prevents its widespread commercial deployment: 
how to generate high-fidelity advertising images that faithfully preserve the visual characteristics of given product images while achieving compelling visual results according to user specifications?

Existing subject-driven generation methods reveal a critical dilemma between visual fidelity and computational efficiency.
On the one hand, tuning-based methods like DreamBooth~\cite{ruiz2023dreambooth} and Textual Inversion~\cite{gal2022image} achieve remarkable fidelity in preserving product details.
However, their ``one-model-per-subject'' paradigm incurs prohibitive training and storage costs, making them impractical for e-commerce platforms managing a large number of products.
On the other hand, tuning-free methods like IP-Adapter~\cite{ye2023ip-adapter} and PhotoMaker~\cite{li2023photomaker} offer the required efficiency and scalability, but they often fail to preserve product details such as the unique textures, shapes, and logos, which are essential for maintaining brand identity in advertising contexts.
This fidelity-efficiency dilemma represents the primary barrier to applying AIGC techniques in advertising image generation.

    \begin{figure}[!t]
    \centering
    \includegraphics[width=0.44\textwidth]{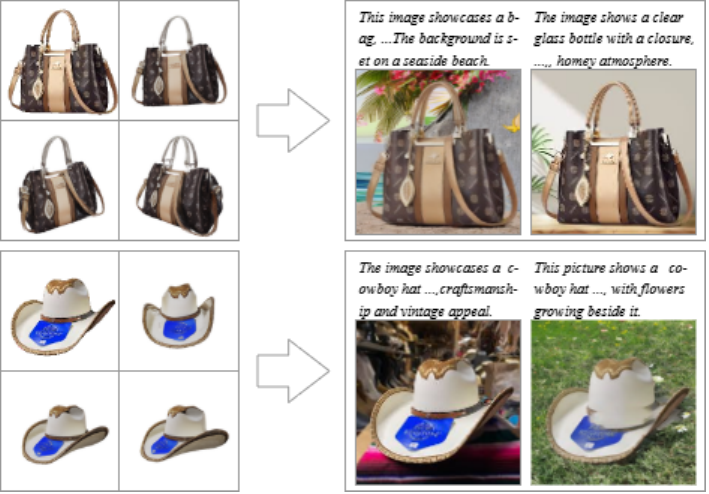}  
    \caption{Examples of generated advertisement images.}
    \label{figure01}
    \vspace{-0.5cm}
    \end{figure}
    
To bridge this fidelity-efficiency gap, we propose \model{}, a novel generation method that achieves high fidelity within a tuning-free paradigm. Its core innovation is a decoupled generation architecture that disentangles the complex synthesis process into two distinct tasks: spatial layout control and identity feature fusion. Spatial control is achieved by injecting a product mask at the U-Net's input, which implicitly guides the model's spatial awareness throughout the generation process, as opposed to applying an explicit mask during feature fusion. By handling these objectives separately, our design philosophy prevents feature entanglement, enabling the model to achieve state-of-the-art product fidelity while maintaining high efficiency (Fig~\ref{figure01}).

However, developing and validating our method requires high-quality training and evaluation data, which is still lacking in this domain.
Therefore, we introduce \data{}, the first large-scale benchmark specifically designed for reference-based advertising image generation.
We first collect 100,000 high-quality advertising images from various e-commerce platforms and brand websites, and then generate corresponding product images and textual scene descriptions.
We further enhance the diversity of dataset with a novel dual data augmentation strategy that combines  multi-view synthesis~\cite{kerbl20233d} and image degradation, which encourages models to learn intrinsic 3D product properties rather than learning a superficial 2D ``copy-paste'' shortcut.
This dataset not only supports the training and evaluation of our proposed method but also provides researchers the basic foundation they need to tackle this important commercial challenge.

The main contributions of this work are as follows:
\begin{itemize}
    \item We introduce AdProd-100K, the first large-scale benchmark for reference-based advertising generation, featuring a novel dual augmentation strategy to foster robust, 3D-aware learning.
    \item We propose RefAdGen, a framework with a novel decoupled design that combines mask-guidance at the U-Net input with an efficient Attention Fusion Module (AFM) to enable high-fidelity product preservation without per-product training.
    \item We conduct extensive evaluation on both \data{} and in-the-wild images to demonstrate the effectiveness of \model{} in maintaining fidelity and generating compelling advertising images based on textual description.
\end{itemize}

\section{Related Work}

We review recent advances in controllable image generation, focusing on two dominant paradigms: fine-tuning-based and tuning-free subject-driven generation. We also discuss how spatial control and identity preservation are handled in related domains such as fashion synthesis and virtual try-on.

\subsection{Subject-driven Generation via Fine-tuning}

Fine-tuning-based methods adapt pretrained diffusion models to specific subjects by updating part or all of the network parameters. Textual Inversion~\cite{gal2022image} represents a subject using learned pseudo-tokens, enabling personalized synthesis through prompt engineering. DreamBooth~\cite{ruiz2023dreambooth} further enhances identity fidelity by directly tuning the UNet with a few subject-specific images. Extensions of this paradigm incorporate optimization acceleration~\cite{Tewel2023KeyLockedRO} and multi-concept blending~\cite{kumari2023multi}, allowing more flexible subject representations. Some recent works also explore fine-tuning in domain-specific contexts such as fashion generation and personalized avatars, achieving high realism under constrained settings. Nevertheless, all these methods rely on the ``one-model-per-subject'' assumption, which demands separate training for each product. This results in high computational cost and storage burden, rendering them unsuitable for commercial platforms where thousands of new products are introduced dynamically.
Recent research has explored fine-tuning in specific domains such as fashion generation and human modeling. For instance, IMAGGarment-1~\cite{shen2025imaggarment} introduces a two-stage pipeline for garment-conditioned generation, while Long-Term TalkingFace~\cite{shen2025long} applies subject-level conditioning for video portrait synthesis. These studies demonstrate strong fidelity but remain limited by their dependence on subject-specific training, motivating our pursuit of tuning-free solutions that offer generalizability and scalability.

\subsection{Tuning-free Controllable Generation}

Tuning-free methods address scalability by eliminating test-time training. These approaches inject subject and structure information via external conditions, enabling flexible generation pipelines. IP-Adapter~\cite{ye2023ip-adapter} introduces a visual encoder and adapter module to inject reference image features into frozen diffusion backbones. PhotoMaker~\cite{li2023photomaker} extends this pipeline for face generation, while other methods explore multi-subject injection~\cite{Subject-Diffusion}. Although efficient, these approaches often compromise on identity preservation, failing to maintain detailed visual attributes such as material, shape, or branding, which are critical for advertising scenarios.
Another line of work focuses on spatial control. ControlNet~\cite{zhang2023adding} and T2I-Adapter~\cite{mou2023t2i} inject structured priors like edge maps or pose skeletons through auxiliary encoders, offering layout consistency in generated results. In the fashion domain, Imagpose~\cite{shen2024imagpose} and Imagdressing-V1~\cite{shen2025imagdressing} leverage pose or mask guidance to enable pose-aligned generation and virtual try-on. While these methods excel at structural fidelity, they do not explicitly address the preservation of instance-level identity, a key requirement in advertising content generation.
Instruction-based editing methods such as InstructPix2Pix~\cite{brooks2022instructpix2pix} guide image synthesis via natural language commands. However, their global editing process often overwrites subject appearance, compromising brand consistency. Our method addresses this by separating spatial layout and identity control. Inspired by IMAGHarmony~\cite{shen2025imagharmony}, we adopt a decoupled design that injects masks into the U-Net for layout control and fuses identity features through a lightweight attention module, enabling high fidelity without fine-tuning.

\begin{figure}[!t]

\includegraphics[width=0.474\textwidth]{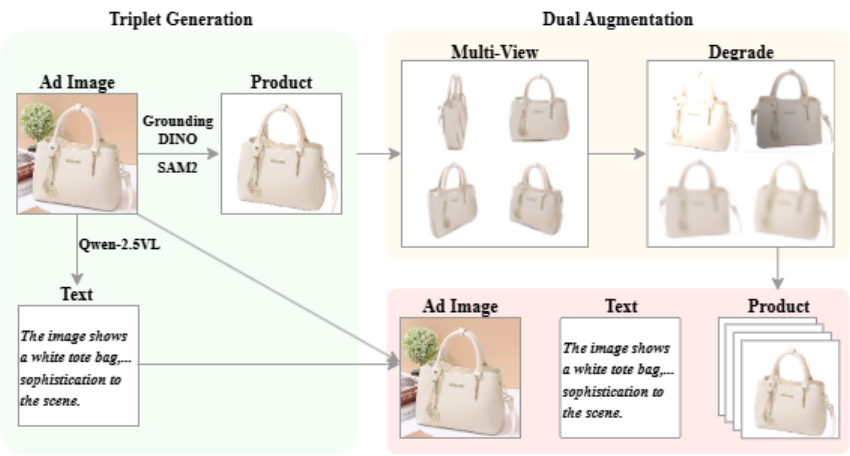}
\caption{The construction pipeline of \data{}. Starting from the advertising images, we first construct triplets by extracting the product images and textual scene description, and then enhance them with dual data augmentation.}
\label{figure02}

\end{figure}

\section{3. Build \data{} with Dual Augmentation}

\begin{figure*}[!t]
\centering
\includegraphics[width=1\textwidth]{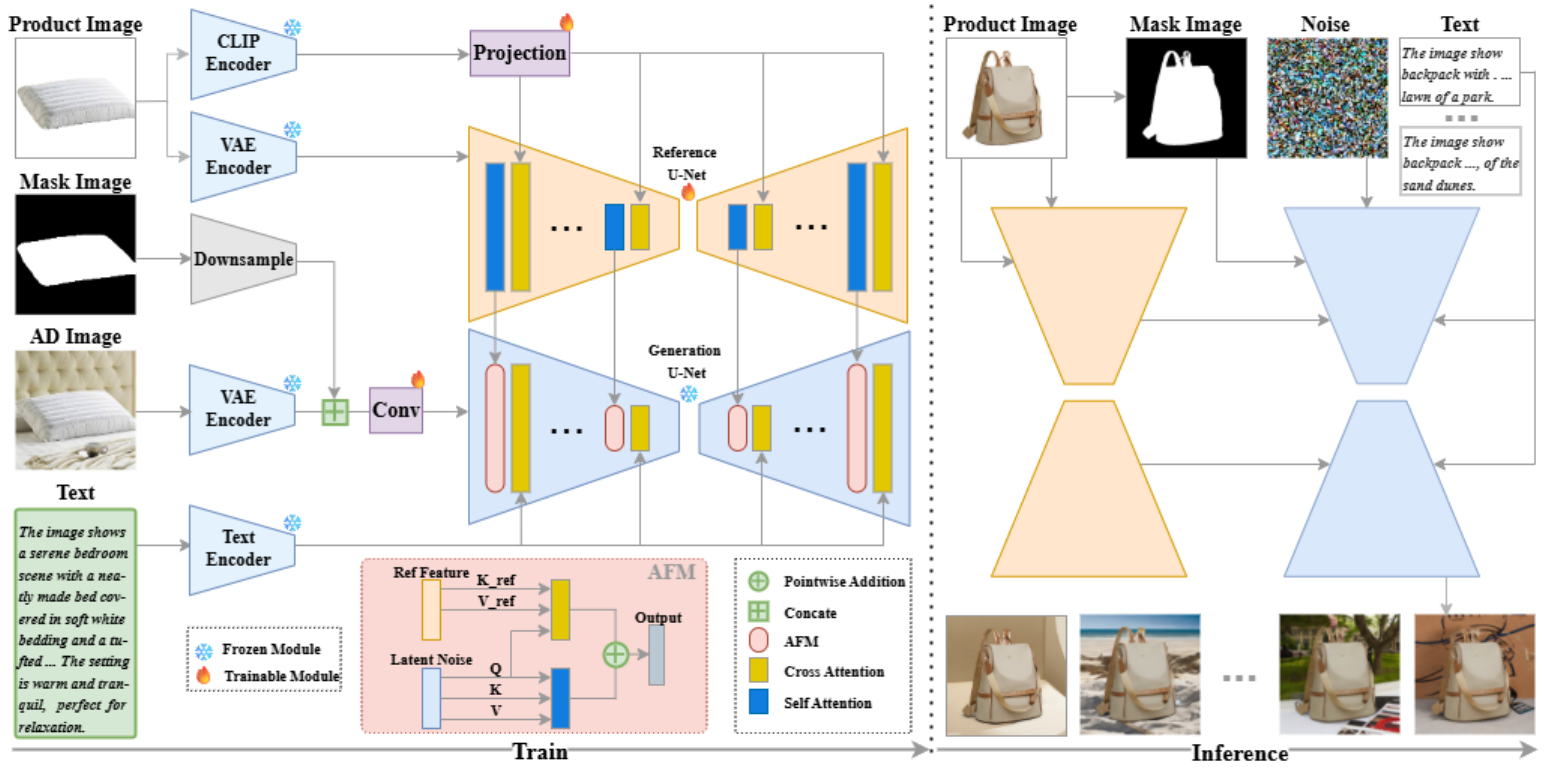}  
\caption{The overall model architecture of RefAdGen, featuring a decoupled dual U-Net design. The Generation U-Net receives the noisy latent and the product mask $\mathbf{M}'$ at its input for spatial control. At each level of the network, the Attention Fusion Module (AFM) fuses identity features from the Reference U-Net with the scene features of the Generation U-Net.}
\vspace{-0.3cm}
\label{figure03}
\end{figure*}

The key to advertising image generation is a large collection of high-quality triplets containing textual scene descriptions, product images, and advertising images.
To achieve this, we constructed \data{}, a large-scale dataset designed for both training and evaluating models.
Figure~\ref{figure02} shows our dataset construction pipeline, which consists of two main stages: quadruple generation and dual augmentation.



\subsection{3.1 Triplet Generation}

Our process begins by collecting over 400,000 images from various e-commerce platforms and brand websites, spanning 30 product categories.
We then used Qwen-2.5VL~\cite{Qwen2.5-VL} to generate the textual scene description, and combined Grounding DINO~\cite{liu2023grounding} and SAM2~\cite{ravi2024sam2} to obtain referenced product images, which are effective and widely used in corresponding tasks.
Throughout this process, we performed a rigorous filtering process to ensure data quality, discarding images with significant watermarks, embedded text, occlusions, or low resolution.
This stage results in 100,000 high-quality triplets, each containing a textual scene description, a product image, and an advertising image.



\subsection{3.2 Dual Augmentation}
Directly training models on these triplets, however, leads to significant limitations and poor generalizability.
First, since the product images are directly segmented from the advertising images, models tend to learn a superficial ``copy-paste'' shortcut.
This results in unrealistic composite images and a lack of diverse viewing angles, as the model merely replicates rather than understands the products.
Second, user-provided product images are inherently of varying quality due to factors like inconsistent lighting, camera shake, and sensor noise.
Without training on such diverse and imperfect data, models struggle to generalize effectively to varied real-world scenarios.



To address these challenges and enhance model robustness, we designed a dual data augmentation strategy: 
\noindent\textbf{(1) Multi-view Image Generation.} This strategy addresses the limitations of a single, fixed camera angle. We leverage 3D Gaussian Splatting~\cite{kerbl20233d} to render novel views from a single product image. Crucially, we recognize that in practical applications, a user's reference image is typically close to their desired output angle. Therefore, our strategy focuses on rendering novel yet similar views, rather than drastically different perspectives. Training with these subtle variations enhances the model's 3D awareness, incentivizing it to learn the product's underlying geometry rather than being restricted to a single, static viewpoint.

\noindent\textbf{(2) Image Degradation.} This strategy simulates real-world imaging variations. By applying degradations like Gaussian noise, shadow variations, and minor geometric warps to the product images, models are forced to distinguish between a product's intrinsic, invariant features (e.g., logo, material) and incidental extrinsic, photographic artifacts. This shifts the model's focus from pixel-level replication to essential feature extraction, enabling it to better handle imperfect input images.

For each original triplet, we generated five augmented variants by modifying the product images.
The mask of each product image is also extracted using SAM2 for model training.
Note that the users can easily generate additional variants using our pipeline.
Table~\ref{table01} provides a detailed variant distribution across different categories.

\begin{table}[!t]
\small
\centering
\setlength{\tabcolsep}{0.4mm}{

\begin{tabular}{l r @{\hspace{2em}} l r @{\hspace{2em}} l r}
\toprule
\textbf{Category} & \textbf{Num} & \textbf{Category} & \textbf{Num} & \textbf{Category} & \textbf{Num} \\
\cmidrule(r){1-2} \cmidrule(lr){3-4} \cmidrule(l){5-6}
Backpack   & 25000 & Eyeshadow  & 13960 & Pens        & 14650 \\
Bench      & 21855 & Fork       & 18115 & Pillows     & 14595 \\
Body wash  & 25000 & Foundation & 10715 & Rugs        & 8760  \\
Bottle     & 25000 & Handbag    & 25000 & Shampoo     & 22990 \\
Car        & 25000 & Hats       & 36740 & Snacks      & 24990 \\
Cell phone & 25000 & Headphones & 22645 & Sneakers    & 24995 \\
Chargers   & 21095 & Kite       & 13030 & Sports ball & 7045  \\
Clothing   & 24995 & Lipstick   & 10585 & Toothbrush  & 7105  \\
Coffee     & 24275 & Motorcycle & 24995 & Umbrella    & 15675 \\
Cup        & 16860 & Notebooks  & 24995 & Wine glass  & 24990 \\
\bottomrule
\end{tabular}
\caption{30 categories and their sample counts. }
\label{table01}}
    \vspace{-0.5cm}
\end{table}

\section{4. \model{}}
In addition to constructing \data{}, we proposed \model{} to generate high-fidelity advertising images without fine-tuning on each input product.
The core idea is to separate the task of scene generation from product injection and then precisely merge them.
Figure~\ref{figure03} illustrates the design of \model{}, which consists of two core components: 1) a dual U-Net backbone, where the Generation U-Net is spatially guided by an input product mask, and 2) an Attention Fusion Module for injecting identity features from the Reference U-Net into the Generation U-Net.

\subsection{4.1 Dual U-Net Architecture}
To address the inherent conflict between diverse scene generation and high-fidelity identity preservation, we adopted a dual U-Net architecture that decouples these competing objectives into specialized processing streams. This architecture leverages two networks derived from the same pre-trained model, ensuring natural feature space compatibility that enables seamless identity-scene fusion while allowing each U-Net to excel in its dedicated task.

\noindent\textbf{Reference U-Net.} This network functions as a dedicated identity feature extractor. Its weights are fully fine-tuned during training to learn and extract the most critical features for identity preservation from the input product image. This design enables it to supply highly relevant and condensed identity information to the generation process.

\noindent\textbf{Generation U-Net.} This network serves as the primary denoising backbone for synthesizing the final advertising images based on the extracted product feature and the user-provided scene description. To incorporate precise spatial guidance, we modified its input to accept a 5-channel tensor, formed by concatenating the standard 4-channel noisy latent $\mathbf{z}_t$ with the 1-channel product mask $\mathbf{M}'$. The first convolutional layer is modified accordingly to accept this extra channel. To preserve the powerful generative prior and achieve parameter-efficient adaptation, most parameters are copied from the pre-trained model and frozen. Only the modified input layer and the projection matrices within our AFM modules are made trainable.

\subsection*{4.2 Attention Fusion Module}
To effectively inject identity features into the scene, we employ an Attention Fusion Module (AFM) that computes self-attention and cross-attention in parallel and then sums their outputs.
Specifically, for a given query feature $\mathbf{Q}$ from the Generation U-Net, the module's final output $\mathbf{O}_{\text{AFM}}$ combines 1) a self-attention term for scene structure, calculated using the Generation U-Net's own key ($\mathbf{K}$) and value ($\mathbf{V}$); and a cross-attention term for identity injection, calculated using the key ($\mathbf{K}_{\text{ref}}$) and value ($\mathbf{V}_{\text{ref}}$) from the Reference U-Net. The entire operation is captured by:
\begin{equation}
    \mathbf{O}_{\text{AFM}} = \text{softmax}\left(\frac{\mathbf{Q}\mathbf{K}^T}{\sqrt{d_k}}\right)\mathbf{V} + \text{softmax}\left(\frac{\mathbf{Q} \mathbf{K}_{\text{ref}}^T}{\sqrt{d_k}}\right) \mathbf{V}_{\text{ref}},
    \label{eq:afm_combined}
\end{equation}
where $d_k$ is the dimension of the key vectors.

\noindent\textbf{Implicit Spatial Control Mechanism.}
To enable 3D-aware generation with novel viewpoints, our method injects masks at the input of Generation U-Net rather than applying rigid feature-level masking at each level.
By providing the mask as an early spatial prior, we allow it to naturally propagate through the network, creating spatially-aware query vectors ($\mathbf{Q}$) that adaptively attend to relevant identity features based on spatial context.
This implicit mechanism delivers two benefits: it maintains a precise spatial layout while encouraging the model to learn intrinsic 3D object representations rather than memorizing static 2D silhouettes.
Consequently, \model{} achieves geometrically coherent synthesis across novel viewpoints.

\subsection{4.3 Training Objective}
We train RefAdGen end-to-end using the standard noise prediction objective from latent diffusion models. The model $\boldsymbol{\epsilon}_\theta$, represented by our AFM-integrated U-Net, is trained to predict the sampled noise $\boldsymbol{\epsilon}$ from the noisy latent $\mathbf{z}_t$. The loss function $\mathcal{L}$ is the mean squared error between the predicted and sampled noise, conditioned on the text prompt $\mathbf{c}_{\text{text}}$, the spatial product mask $\mathbf{M}$, and the feature $\mathbf{z}_{\text{ref}}$, which is encoded from the product image by our Reference U-Net.
\begin{equation}
\scalebox{0.9}{$
    \mathcal{L} = \mathbb{E}_{\mathbf{z}_0, \mathbf{c}_{\text{text}}, \mathbf{z}_{\text{ref}}, \mathbf{M}, \boldsymbol{\epsilon}, t} \left[ \left\| \boldsymbol{\epsilon} - \boldsymbol{\epsilon}_\theta\left(\text{concat}(\mathbf{z}_t, \mathbf{M}'), t, \mathbf{c}_{\text{text}}, \mathbf{z}_{\text{ref}}\right) \right\|_2^2 \right].
$}
    \label{eq:loss_final}
\end{equation}

\begin{figure*}[!t]
\centering
\includegraphics[width=\textwidth]{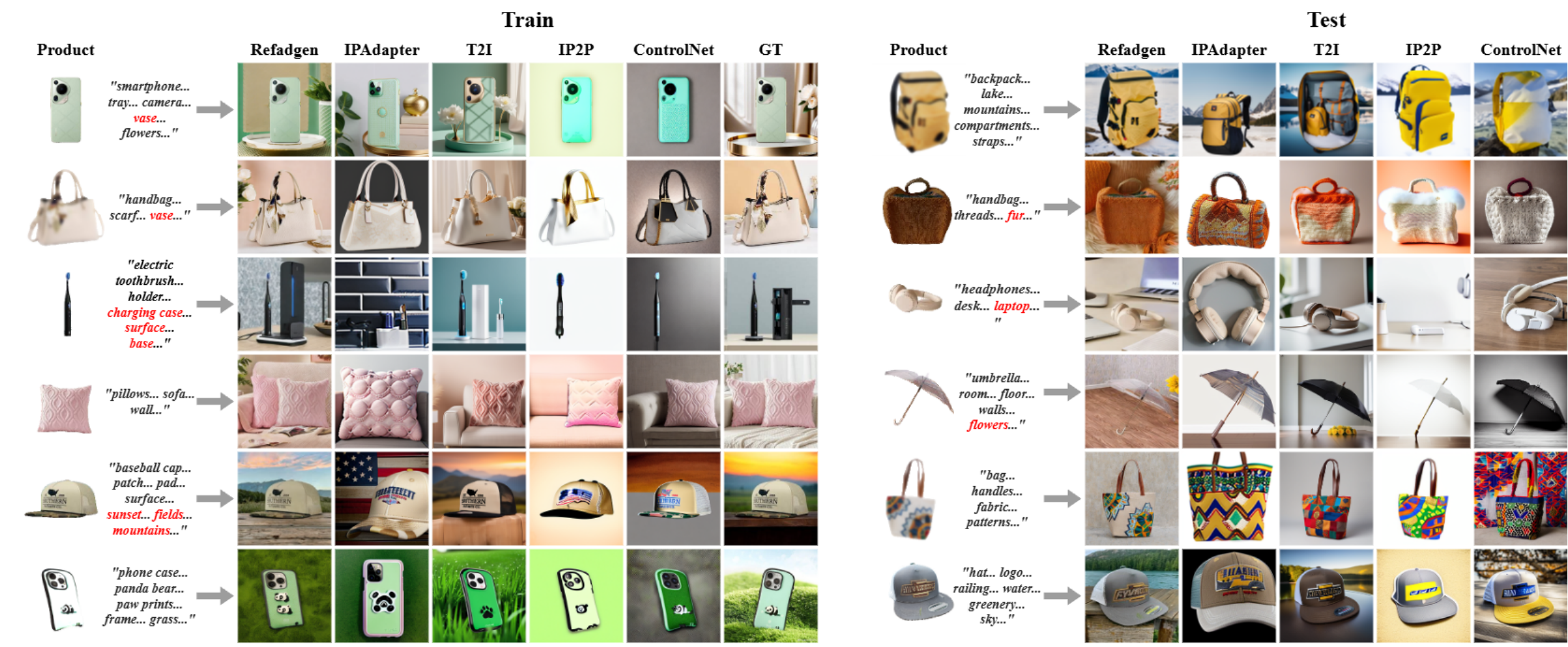}
\vspace{-0.5cm}
\caption{Qualitative comparisons on \data{}. Prompts are simplified for clarity. Both the training samples on the left and the test samples on the right showcase the consistent advantages of RefAdGen in identity consistency, scene realism, and overall aesthetic quality.}
\label{figure04}
    \vspace{-0.3cm}
\end{figure*}

\begin{figure}[t]
\centering
\includegraphics[width=\columnwidth]{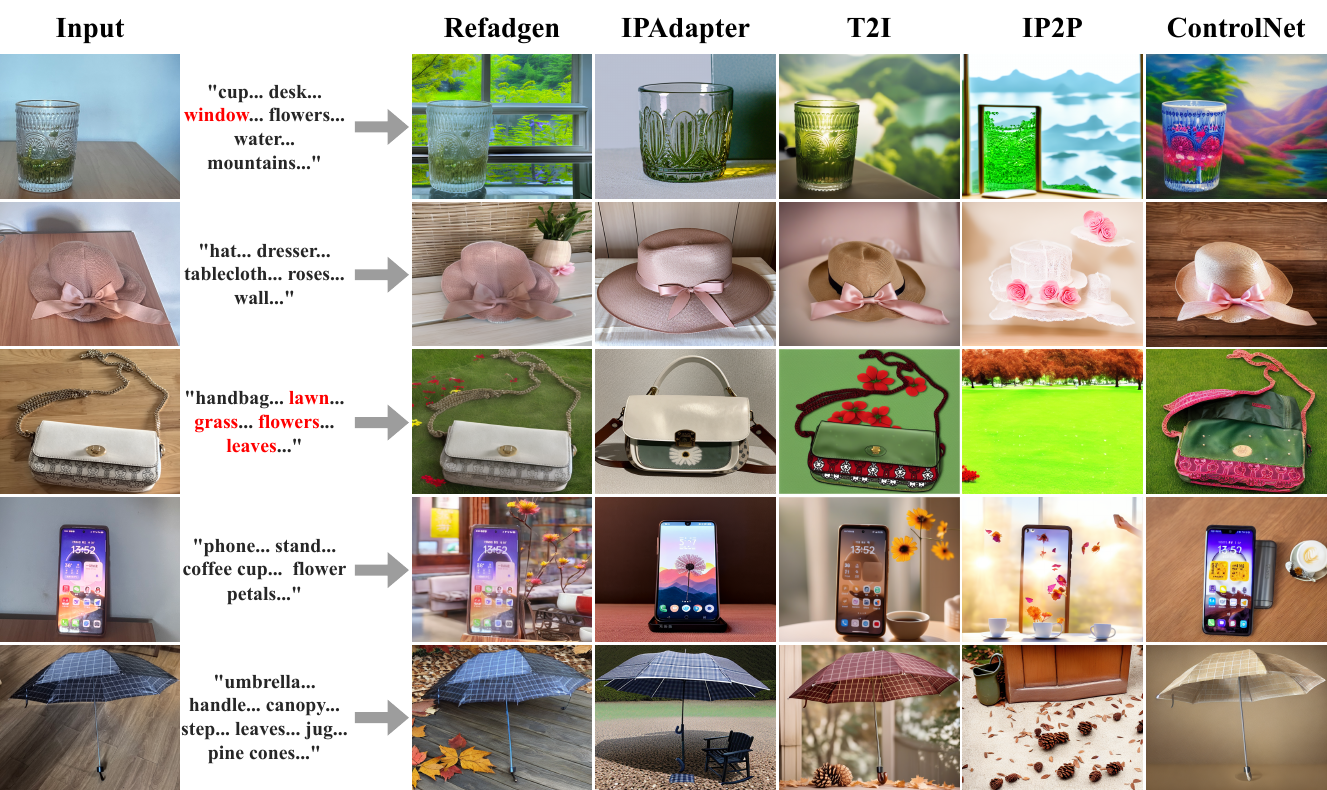}
\caption{Qualitative comparison on in-the-wild images. The inputs are real-world photos captured using mobile phones, from which our pipeline automatically segments the product image and generates advertising images using \model{}.}
    \label{figure05}
        \vspace{-0.3cm}
\end{figure}

\section{5. Experiments}

This section systematically validates our proposed framework, RefAdGen, through a series of comprehensive experiments conducted on our \data{}.
In addition to numerical evaluation, we further tested our method on a diverse set of in-the-wild photographs captured with mobile phones, confirming its robustness and broad applicability.
A user study is also conducted to assess human perception of our generated results.

\subsection{5.1 Experimental Setup}
\begin{table}[!t]

    \centering
    \small

    \begin{tabular}{llll}
        \toprule
        Hyperparameter      & Value         & Hyperparameter      & Value \\
        \midrule
        Optimizer           & AdamW         & Weight Decay        & 0.01 \\
        Batch Size         & 3             & Learning Rate       & $1 \times 10^{-5}$ \\
        Noise Offset       & 0.05          & LR Scheduler        & Linear \\
        Training Epochs    & 8             & Warmup Steps   & 500 \\
        \bottomrule
    \end{tabular}
    \caption{Hyperparameter configuration for the model.}
        \label{table02}
        \vspace{-0.6cm}
\end{table}

\noindent\textbf{Dataset and Metrics}. Our experiments were conducted on the \data{} dataset, using a 9:1 training-to-testing split.
We evaluated performance on five metrics across several key dimensions: CLIP-Score~\cite{hessel2021clipscore} for text adherence, FID~\cite{Seitzer2020FID} for realism, ImageReward~\cite{xu2023imagereward} for human preference, and MP-LPIPS~\cite{chen2024magic} and LPIPS~\cite{zhang2018perceptual} for perceptual similarity and identity preservation.
Beyond qualitative analysis, we also conducted qualitative comparison on in-the-wild images to comprehensively evaluate model performance and generalization.

\noindent\textbf{Baselines}. We compared RefAdGen against state-of-the-art methods representing diverse control strategies: IP-Adapter~\cite{ye2023ip-adapter} that uses identify as reference, ControlNet~\cite{zhang2023adding} and T2I-Adapter~\cite{mou2023t2i} that uses structure information as reference, and Instruct-Pix2Pix (IP2P)~\cite{brooks2022instructpix2pix} that performs instruction-guided image editing. To ensure a fair comparison, all baselines and our model are built upon the Stable Diffusion v1.5 backbone.
\begin{table*}[!t]
    \centering
    \sisetup{detect-weight, mode=text}
    \renewcommand{\arraystretch}{1.2}
    \begin{tabular*}{\linewidth}{@{\extracolsep{\fill}} l S[table-format=2.4] S[table-format=2.4] S[table-format=-1.4] S[table-format=1.4] S[table-format=1.4] @{}}
        \toprule
        \textbf{Model} & {\textbf{CLIP Score}$\uparrow$} & {\textbf{FID}$\downarrow$} & {\textbf{ImageReward}$\uparrow$} & {\textbf{MP-LPIPS}$\downarrow$} & {\textbf{LPIPS}$\downarrow$} \\
        \midrule
        IP-Adapter      & 32.6308 & 62.6666 & -0.2572 & 0.3974 & 0.7159 \\
        T2I             & 34.2737 & 59.1770 &  0.1777 & 0.3382 & 0.6063 \\
        IP2P            & 32.4433 & 64.1141 & -0.5842 & 0.3517 & 0.6336 \\
        ControlNet      & 33.0226 & 57.1835 & -0.3668 & 0.3748 & 0.6578 \\
        \midrule
        RefAdGen (Ours) & \bfseries 34.5106 & \bfseries 50.5843 & \bfseries 0.2391 & \bfseries 0.2612 & \bfseries 0.5487 \\
        \bottomrule
    \end{tabular*}
    \caption{
        Performance comparison of our model (RefAdGen) against several baselines on five key metrics.
        The arrow indicates whether a higher value ($\uparrow$) or a lower value ($\downarrow$) is better.
        The \textbf{best} result in each column is highlighted in bold.
    }
    \label{table03}
        \vspace{-0.5cm}
\end{table*}

\noindent\textbf{Implementation Details.} Our model was trained on two NVIDIA 5090D GPUs. All input images were resized to $512 \times 640$. Detailed hyperparameter settings are provided in Table~\ref{table02}.

\subsection{5.2 Main Results and Analysis}

\noindent\textbf{Quantitative Comparison}.
The quantitative results in Table~\ref{table03} demonstrate the effectiveness of our \model{}.
A central challenge is maintaining perceptual similarity while ensuring faithful text adherence.
Methods like T2I perform well in text adherence (34.27 CLIP-Score) but  exhibit significant weaknesses at preserving the fine-grained details (0.3382 MP-LPIPS).
Conversely, identity-reference methods like IP-Adapter compromise on text adherence and overall realism in favor of perceptual similarity.
In contrast, our method achieves state-of-the-art performance in both dimensions, scoring highest on CLIP-Score (34.5106), MP-LPIPS (0.2612), and LPIPS (0.5487).
This demonstrates its robust ability to integrate the reference product seamlessly into a novel scene described by the text prompt.

In addition, this high-fidelity generation does not compromise overall image quality or aesthetic appeal.
\model{} achieves the best realism with an FID score of 50.58, which indicates that our generated scenes are not only accurate but also visually harmonious. 
Our model also scores highest on ImageReward (0.2391), which suggests that our results possess strong commercial appeal for advertising applications.

\noindent\textbf{Qualitative Comparison}.
To comprehensively evaluate \model{}, we also conducted qualitative analysis on our \data{} benchmark (Figure~\ref{figure04}).
The results on training dataset (left) verify that \model{} generates novel images rather than simply memorizing the ground truth images, which confirms the absence of severe overfitting.
The results on the unseen test dataset (right) demostrate that \model{} maintains high fidelity for new items, proving its strong generalization capabilities.

To further assess robustness and practical utility, we evaluated model performance on challenging in-the-wild images (Figure~\ref{figure05}).
For these mobile phone-captured images, we first extracted product images using the same method empolyed in our data generation pipeline, and then generated advertising images using \model{}.
The results show that \model{} not only handles this domain shift successfully but also exhibits remarkable lighting adaptability, endering objects with physically plausible illumination that seamlessly integrates with the target scene context.
These evaluation results confirm the effectiveness and robustness of our method.

\begin{figure}[!t]
\centering
\includegraphics[width=\columnwidth]{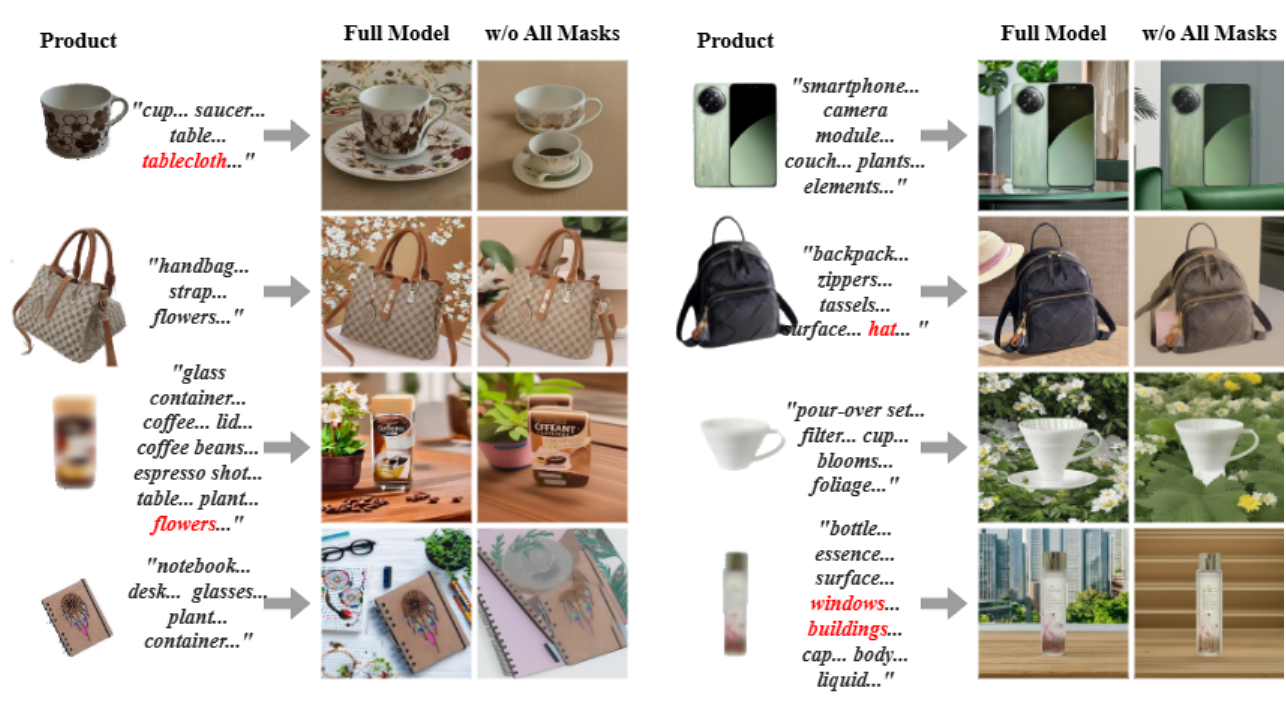}
\vspace{-1em}
\caption{Qualitative ablation analysis of architectural components. The Full Model better preserves details, while the ablation variant without U-Net input masks exhibits degradation in object placement precision and scene coherence.
}
    \label{figure06}
    \vspace{-0.3cm}
\end{figure}

\begin{figure}[!t]
\centering
\includegraphics[width=\columnwidth]{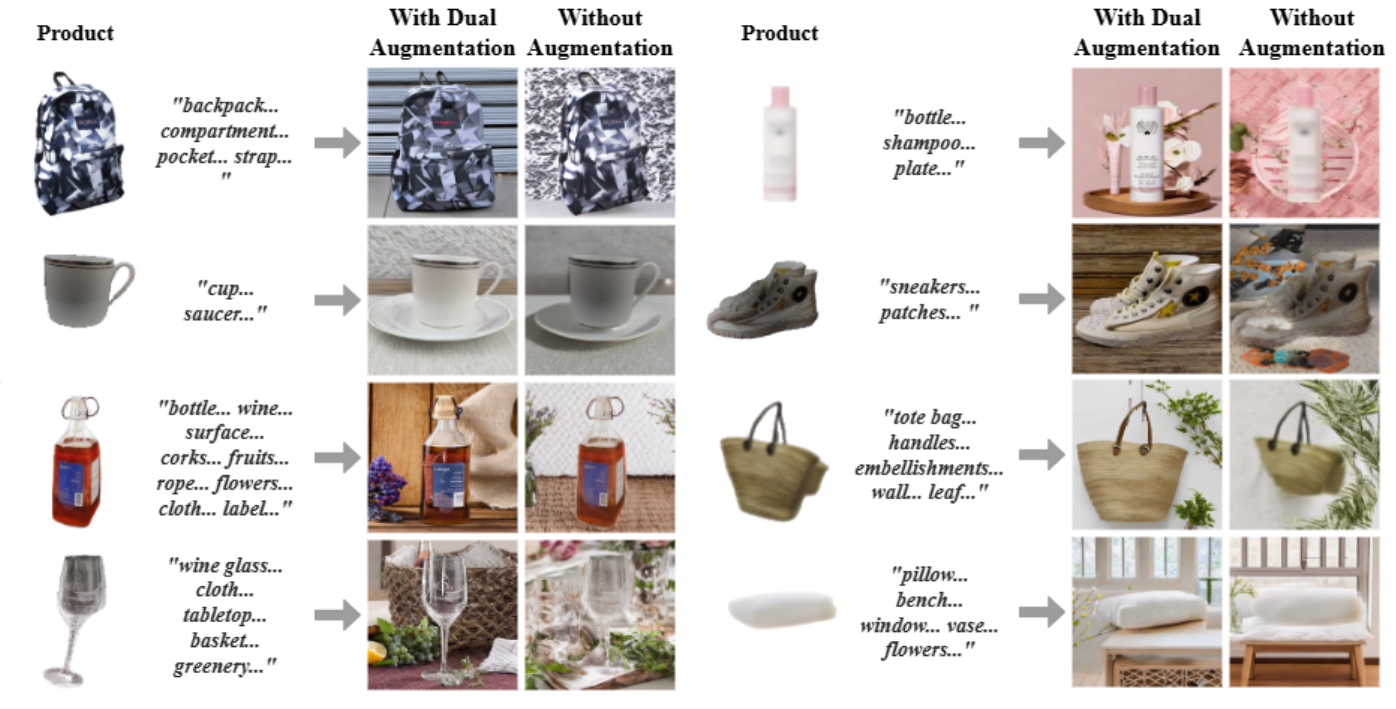}
    \caption{Qualitative ablation analysis of the Dual Augmentation strategy. With augmentation, the generation results exhibit substantially enhanced visual clarity and improved robustness compared to the non-augmented baseline.}
    \label{figure07}
     \vspace{-0.5cm}
\end{figure}

\begin{figure*}[!t]
\centering
\includegraphics[width=1\textwidth]{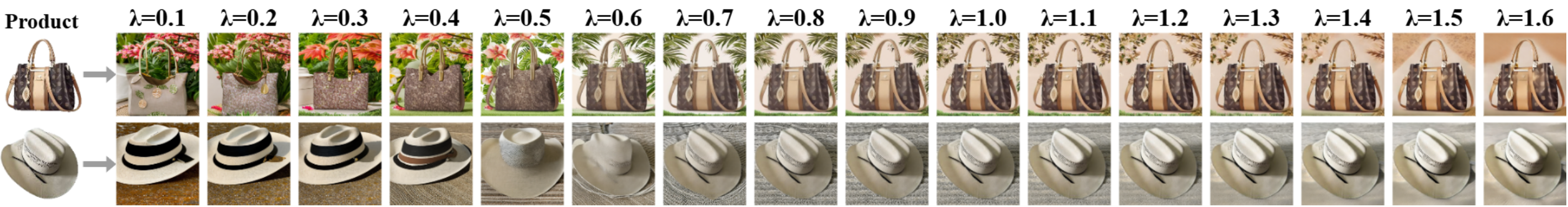}  
\caption{Image generation fidelity improves markedly and remains consistently high for all $\lambda$ values above 0.7.}
\label{figure08}
 \vspace{-0.2cm}
\end{figure*}

\subsection{5.3 Ablation Studies}

To systematically dissect the contributions of our design choices, we conducted a comprehensive ablation study covering two key aspects: model architectural components and data augmentation strategies.
\begin{table*}[t]
    \centering
    \sisetup{detect-weight, mode=text}
    \renewcommand{\arraystretch}{1.2} 
    \begin{tabular*}{\linewidth}{@{\extracolsep{\fill}} l S[table-format=2.4] S[table-format=2.4] S[table-format=-1.4] S[table-format=1.4] S[table-format=1.4] @{}}
        \toprule
        \textbf{Configuration} & {\textbf{CLIP Score}$\uparrow$} & {\textbf{FID}$\downarrow$} & {\textbf{ImageReward}$\uparrow$} & {\textbf{MP-LPIPS}$\downarrow$} & {\textbf{LPIPS}$\downarrow$} \\
        \midrule
        \bfseries Full Model (Ours) & \bfseries 34.5106 & \bfseries 50.5843 & \bfseries 0.2391 & \bfseries 0.2612 & \bfseries 0.5487 \\
        \midrule
        w/o Masks               & 33.2415 & 55.0244 & -0.2293 & 0.3638 & 0.6602 \\
        w/o Dual Augmentation   & 32.7952 & 68.7243 & -0.3554 & 0.3014 & 0.6064 \\
        \bottomrule
    \end{tabular*}
    \caption{
        Ablation study of our framework's core components and data strategy. ``Full Model" represents our complete design, while each subsequent row ablates one key element. The results highlight the critical contributions of each component. The \textbf{best} score in each column is highlighted in bold.
    }
    \label{tab:ablation_compact}
        \vspace{-0.5cm}
\end{table*}

\begin{table}[ht]
  \centering
  \setlength{\tabcolsep}{1.3mm}{
  \begin{tabular}{lccccc}
    \toprule
    & \model{} & ControlNet & IP2P & IP-Adapter & T2I \\
    \midrule
    J2b & \bfseries 38.70 & 23.60 & 8.40 & 0.90 & 28.40 \\
    G2R & \bfseries 90.70 & 77.30 & 47.30 & 6.70 & 78.00 \\
    \bottomrule
  \end{tabular}}
  \caption{Results of our user study comparing RefAdGen with baseline methods. The \textbf{best} score is highlighted in bold.}
  \label{table05}
      \vspace{-0.5cm}
\end{table}

\noindent\textbf{Analysis of Model Architectural Components}.
We first analyze the contributions of our product mask component.
As shown in Table~\ref{tab:ablation_compact}, removing mask guidance from the generation U-Net input results in a significant drop in performance, particularly for realism (FID) and identity preservation (MP-LPIPS).
This provides empirical justification for our model design: injecting the mask at the input layer provides essential spatial bias, enabling the network to learn selective application of identity features.
Without this spatial guidance, the model fails to perform implicit feature gating, resulting in feature bleeding and consequent degradation in generation quality.

\noindent\textbf{The Foundational Role of Dual Augmentation}. Next, we ablate our data generation method.
As shown in Table~\ref{tab:ablation_compact}, removing dual data augmentation results in a catastrophic drop in performance, with the FID score increasing dramatically from 50.58 to 68.72.
This suggests that without our augmentation strategy, the model cannot effectively generalize to create realistic, well-integrated scenes and instead resorts to superficial ``copy-paste" behavior from training data.

Figure~\ref{figure07} provides the evidence supporting these quantitative findings.
Across all test cases, the model trained with our augmentation produces clearer, more realistic, and more robust results.
It also successfully handles varied lighting, generates novel viewpoints (e.g., the cup and sneakers), and reconstructs details from imperfect inputs.
In contrast, the model trained without augmentation learns a superficial ``copy-paste" of the input.

\subsection{5.4 User Study}
To assess human perception, we also conducted a subjective study with 30 volunteers using two protocols. For Realism Judgment (G2R), participants were shown individual images and asked to distinguish whether they were real photographs or AI-generated, thereby measuring perceived authenticity. For Preference Selection (J2b), we presented the image from our method alongside the images from all baselines simultaneously, and asked users to select the single best result based on realism and aesthetic quality.
As shown in Table~\ref{table05}, our method significantly outperforms all baselines on both metrics. This indicates that images generated by RefAdGen are not only highly realistic but are also more aesthetically favored by users.

\subsection{5.5 Further Analysis and Discussion}

To provide a more comprehensive understanding of our model's performance, we conducted additional analyses.

\begin{figure}[!t]
\hspace{-1em}
\includegraphics[width=0.486\textwidth]{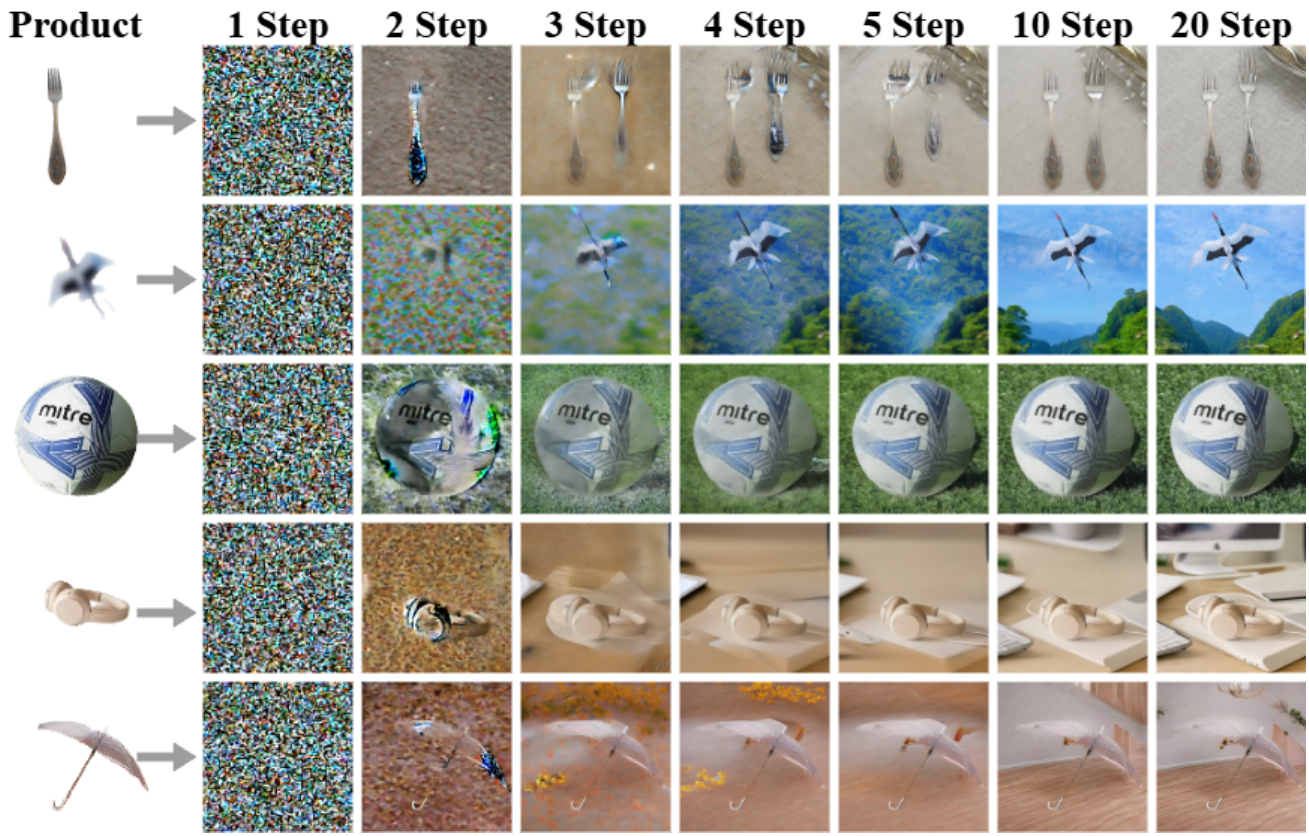}  
\caption{Visualization of the denoising process.}
\label{figure09}
    \vspace{-0.5cm}
\end{figure}

\noindent\textbf{Robustness to Fusion Strength}. 
Balancing multiple guidance signals often requires careful hyperparameter calibration.
We thus tested our identity fusion's sensitivity to a trade-off parameter, $\lambda$, in the fusion mechanism ($\mathbf{O}_{\text{AFM}} = \mathbf{O}_{\text{self}} + \lambda \cdot \mathbf{O}_{\text{cross}}$).
As shown in Figure~\ref{figure08}, our method is remarkably stable, maintaining consistently high-quality results where $\lambda \ge 0.7$.
This robustness stems from our decoupled architecture.
The spatial guidance provided by the input mask enables the model to apply identity features only within the intended foreground, mitigating feature conflicts that would otherwise necessitate a trade-off.
This spatial disentanglement eliminates the criticality of $\lambda$, allowing us to set a default value of 1.0.

\noindent\textbf{Analysis of Generation Process}.
The progressive denoising process, illustrated in Figure~\ref{figure09}, demonstrates the computational efficiency of our method.
The overall structure emerges within the first 3 steps, with fine-grained details being refined from step 10.
High-quality results are achieved by approximately step 20, demonstrating a good balance between computational efficiency and output quality.

\section{6. Conclusion and Future Work}

This paper addresses the fidelity-efficiency trade-off in reference-based generation. We introduce \data{}, a large-scale benchmark for this task, and \model{}, a tuning-free framework with a decoupled design. It achieves state-of-the-art identity preservation through the synergy of its mask-guidance at the U-Net input and an efficient Attention Fusion Module (AFM).

\noindent \textbf{Future Work}. We aim to extend our framework to video ad generation and to build a closed-loop system that optimizes content directly for business metrics, such as conversion rates.

\bibliography{aaai2026}

\end{document}